\documentstyle[prl,multicol,aps]{revtex} % within 4 PRL pages
\tighten

\begin{document}
\draft

\title
{The Hubbard quantum wire}

\author
{You-Quan Li$^{1,2}$ and Christian Gruber$^{1}$ }

\address
{
${}^{1}$ Institut de Physique Th\'{e}orique, 
\'{E}cole Polytechnique F\'{e}d\'{e}rale de Lausanne, 
CH-1015 Lausanne \\
${}^{2}$ Zhejiang Institute of Modern Physics,
Zhejiang University, Hangzhou 310027, P. R. China 
}

\date{ Received September 9, 1997 }

\maketitle

\begin{abstract}

By introducing a boundary condition for the quantum wire,
the Hubbard model is solved exactly by means of Bethe ansatz. 
The wave function for the bounded state is clearly defined, 
and the secular equation for the spectrum is exactly obtained.
The ground state and low-lying excited states are studied 
in the thermodynamic limit. The ground-state energy
in the strong coupling limit is obtained 
explicitly, and compared with the results of periodic boundary
condition.
\end{abstract}


\begin{multicols}{2}

It is known that the behavior of one-dimensional electronic 
system differs in many aspects from that of two-dimensional
and three-dimensional systems \cite{Soly}. 
The one-dimensional Hubbard model provides 
the opportunity to study correlation effects.
There has been many discussions on Hubbard model 
since the Lieb-Wu solution was found \cite{LiebWu}. 
This solution is an exact result with periodic boundary 
condition which can be thought of as on a ring. 
Owning to the dramatic achievement in nanotechnology 
in recent years,
the boundary effects of a sample will become more important.
The Hubbard model on an open chain with completely 
confined ends was considered \cite{Schulz}.
By adding some boundary field at the ends, the model
was considered \cite{ZSW} and discussed \cite{BedFrahm}.
The models \cite{Schulz,ZSW,BedFrahm} were easier and simpler
as the wave function outside the wire is null.
In present letter we study a more realistic model 
for quantum wires where the possibility of a nonvanishing 
wave function outside is taken into account. 
Using Bethe ansatz and some 
other methods in the literature of integrable models, we obtain
an exact result for the model hamiltonian. The ground state and
the low-lying excitations are studied in the thermodynamic limit.
The ground state energy in the strong coupling
is shifted from the one for periodic boundary condition in a small amount.
Both the spinon excitation and the holon-antiholon excitation
are gapless. 

We consider a quantum wire of length $L$ described by the
hamiltonian
\begin{eqnarray}
H = \sum^{\infty}_{\stackrel{i=-\infty}{a} }
        \left[-t(C^{+}_{i a}C_{i+1 a} + C^{+}_{i+1 a}C_{i a})
               + \frac{U}{2} n_{i a}n_{i -a}
         \right] \nonumber \\       
  + \sum^{ -L/2 -1}_{ \stackrel{i=-\infty}{a} }V_L n_{i a} 
  + \sum^{\infty}_{ \stackrel{i=L/2+1}{a} }V_R n_{i a}, 
\label{eq:Hamiltonian}
\end{eqnarray}      
where $C_{i a }$ is the operator annihilating an electron  
with spin component $a$ on site $i$, and 
$n_{i a}:=C^{+}_{i a}C_{i a}$ the local-number operator of 
electrons.
Clearly, it is a Hubbard hamiltonian added with some terms
which are defined outside of the wire. So the present model
is different from either Hubbard ring \cite{LiebWu}
or the Hubbard chain with reflection ends \cite{Schulz}
or with boundary fields \cite{ZSW,BedFrahm}.
The $V_L$ and $V_R$ in (\ref{eq:Hamiltonian})
represent the kinetic energies which an electron must lose 
if it escapes out of the wire from left end or right 
end respectively. Instead of periodic or completely
confined boundary condition, 
we will treat in this model with a more physical 
boundary condition.  It is convenient to consider
the states that span a Hilbert space of $N$-particles
\[
   |\psi > =\sum_{\{a_i \}, \{x_i \} }
          \psi_{a_1,\cdots a_N }(x_1,\cdots, x_N)    
          C^{+}_{x_1 a_1} \cdots C^{+}_{x_N a_N}|0>.
\]
The eigenvalue problem 
$ H |\psi> = E |\psi>$
becomes an $N$-particle quantum mechanical problem
with the first quantized hamiltonian,
\begin{equation}
H = \sum_{j=1}^{N}
      \left[ - t\Delta_j + V(x_j) 
        \right] 
    + U\sum_{i<j} \delta(x_i , x_j ),    
\label{eq:Schoperator}
\end{equation}
where 
$
\Delta_j \psi := 
   \psi (\cdots, x_j + 1, \cdots )
     +\psi (\cdots, x_j - 1, \cdots ),
$\,\,
$V(x_j)=V_L\theta(-x_j-L/2-1)+V_R\theta(x_j-L/2-1)$,
$\theta(x)$  the step function.
The continuity limit of eq.(\ref{eq:Schoperator})  
concerns a equation in ref.\cite{Li}.

The number operator of electrons in the wire is defined as
$\hat{N}=\sum^{L/2}_{i=-L/2}C^{+}_{ia} C_{ia}$, and that in
the left side or the right side is given by
$\hat{N}_L=\sum^{-L/2-1}_{i=-\infty}C^{+}_{ia} C_{ia}$ or
$\hat{N}_R=\sum^{\infty}_{i=L/2+1}C^{+}_{ia} C_{ia}$
respectively (summation also runs over spins $a=1/2,-1/2$).
The total number operator  
$\hat{N}_{t}=\hat{N}_L +\hat{N}_R +\hat{N}$ 
always gives 
$\hat{N}_t |\psi>=N|\psi>$ in the Hilbert space of
$N$-particles. 
The boundary condition 
which will be used to determine the spectrum reads

\begin{equation}
\lim_{ i\rightarrow\pm\infty}C^{+}_{ia}C_{ia}|\psi> = 0.  
\label{eq:BC}
\end{equation}

The Schr\"{o}dinger operator (\ref{eq:Schoperator}) is 
invariant under any permutation of $S_N$, but is not 
invariant under translation. Thus the total momentum of the 
system is not conserved, and the reflective waves must be 
taking into account. The wave function of Bethe ansatz form
in the region  
$ x\in{\cal C}( Q ):=\{x | -L/2 < x_{Q1}<\cdots<x_{QN}< L/2 \}$
reads
\begin{equation}
\psi_a (x)=\sum_{P \in {\cal W}_B } 
           A_a (P, \, Q ) 
           e^{i(Pk|Qx)},
          \label{eq:BAW}
\end{equation}
where $x:=(x_1, x_2,\cdots, x_N)$ with $x_j\in Z\!\!\! Z$; 
$a:=(a_{Q1}$, $a_{Q2},\cdots,a_{QN})$, $a_j$ 
stands for the spin component of the $j$th particle; 
$Pk$ (or $Qx$) is the image of a given 
$k:=(k_1, k_2,\cdots, k_N )$ (or $x$) by 
a mapping $P\in {\cal W}_B $ (or $Q\in{\cal W}_A$); 
$(Pk | Qx) = \sum_{j=1}^{N} (Pk)_j (Qx)_j$. 
The coefficients $ A(P, Q)$ 
are functionals on 
${\cal W}_{B} \otimes {\cal W}_{A}$.
One may notice that the sum runs over the Weyl group
\cite{Gilmore} of the Lie algebra 
$B_N $ (denoted by ${\cal W}_{B}$)
but the wave function is defined on various Weyl chambers
corresponding to the Weyl group of the Lie algebra $A_{N-1}$
(denoted by ${\cal W}_{A}$). 

Any element of the Weyl group ${\cal W}_{B}$
can be expressed
as a product of the neighboring
interchanges,
$\sigma^j : ( \cdots, z_j, z_{j+1}, \cdots )
  \mapsto ( \cdots, z_{j+1}, z_j, \cdots )$
and mirror reflection 
$\tau^1 : (z_1, z_2, \cdots ) 
           \mapsto (- z_1, z_2, \cdots )$,
where $z_j$ denotes either $x_j, a_j, k_j$
or their image by a mapping.
The requirement of anti-symmetry is
$(\sigma^j \psi )_a (x) = - \psi_a (x)$,
which gives 
\[
A(P; \, \sigma^j Q) = 
- {\cal P}^{Qj,Q(j+1)} A(\sigma^j P; \, Q),
\]
where the spin labels are omitted and
${\cal P}^{Qj,Q(j+1)}$ is the spinor representation 
of the permutation
$\sigma^j $.
The Scattering matrix (S-matrix) which relates 
the coefficients $A's$ between distinct regions in the 
configuration space of $N$ electrons reads
\begin{equation}
S^{Qj,Q(j+1)}=
   \frac{\sin(Pk)_j -\sin(Pk)_{j+1} + ic{\cal P}^{Qj,Q(j+1)} }
        {\sin(Pk)_j -\sin(Pk)_{j+1} + ic} 
\label{eq:Smatrix}
\end{equation}
where $2c=U/t$.
Eq.(\ref{eq:Smatrix}) is the same as that in the Lieb-Wu 
solution \cite{LiebWu}.
The coefficients $A's$
in any region are determined up to an overall factor 
by the $\check{S}^{Qj,Q(j+1)}:= 
  - {\cal P}^{Qj,Q(j+1)} S^{Qj,Q(j+1)}$ 
and the $R_L$ (or $R_R$),  i.e., 
$A(\sigma^j P; Q) = \check{S}^{Qj,Q(j+1)} A(P; Q)$,
$A(\tau^1 P; Q)=R_L[ (Pk)_1] A(P; Q)$,
$A(\tau^N P; Q)=R_R[ (Pk)_N] A(P; Q)$.
The reflection matrices  $R_L$ and $R_R$ are solved from the 
Schr\"{o}dinger equation near the ends of the wire by
taking into account the boundary condition (\ref{eq:BC}),
consequently,
\begin{eqnarray}
R_L (k) = - e^{-ikL -i\theta_L (k)}, 
  \nonumber \\
R_R (k) = - e^{ikL +i\theta_R (k)},
\label{eq:BSmatrix}
\end{eqnarray}
where
$ i\theta(k) = 
    \ln[(e^{\kappa + ik} - 1)
      /(e^{\kappa - ik} - 1)] $
with $2t(\cosh\kappa -\cos k)=V_L$ or $V_R$
in $\theta_L$ or $\theta_R$.

One may notice that the condition (\ref{eq:BC}) does not
give any constraints directly on $\psi$ 
but gives that on $\psi^L$
and $\psi^R$. In deriving (\ref{eq:BSmatrix})
we have considered the wave functions outside of the wire,
which actually vanishes in \cite{Schulz} but does not 
in present case.
The wave function in the region
$x_{Q1} < -L/2 \leq x_{Q2} < \cdots < x_{QN} \leq L/2$ 
takes  
\[
 \psi_a^{L} (x) = \sum_{j=1}^{N}\sum_{\sigma' \in {\cal W}' }
     A^{L}_a (\sigma' q'_{j}, Q) e^{ \kappa_{j} x_{Q1} }
       e^{ i (\sigma' q'_{j} k |Q'x ) }
\]
where
$Q'x =(0, x_{Q2},\cdots, x_{QN})$ and
$\kappa_{j} > 0$ with 
$ 2t(\cosh \kappa_j - \cos k_j)=V_L$;
in the region
$ -L/2< x_{Q1} < x_{Q2} < \cdots < x_{Q(N-1)} \leq
L/2< x_{QN} $,
\[
  \psi_a^{R}(x) = \sum_{j=1}^{N}\sum_{\sigma'' \in {\cal W}''}
    A^{R}_a (\sigma'' q''_j,Q) e^{ - \kappa_j x_{QN} }
      e^{i(\sigma'' q''_j k |Q''x )}
\]
here $Q''x = (x_{Q1},\cdots, x_{Q(N-1)}, 0)$ and 
$\kappa_j > 0$ with 
$2t(\cosh\kappa_j - \cos k_j)=V_R $.
In the above, some notation conventions are adopted , i.e.,
two subgroups of the Weyl group of $B_N$,
$ {\cal W}':=\{\sigma^2 , \sigma^3 , 
    \cdots,\sigma^{N-1},\tau^N \,\}$
and
${\cal W}'':=\{\sigma^1 , \sigma^2 , \cdots, \sigma^{N-2},
   \sigma^{N-1}\tau^N \sigma^{N-1} \};$
two particular cycles 
$ q''_j: (x_1 , \cdots , x_j, \cdots , x_N )$ 
$ \mapsto$
$( x_1 , \cdots , x_{j-1}, x_{j+1}, \cdots , x_{N}, x_j )$
and
$q'_j: (x_1 , \cdots , x_j, \cdots , x_N )$ 
$\mapsto$
$(x_j , x_1 , \cdots , x_{j-1}, x_{j+1}, \cdots , x_N)$.

As there are several identities in the Weyl groups,
several consistency relations must be checked for
the above solution.
The Yang-Baxter equations \cite{Yang} arising from  both
$A(\sigma^j \sigma^{j+1} \sigma^j P; Q) \,=\,
A(\sigma^{j+1} \sigma^j \sigma^{j+1} P; Q)$ and
$A(P; \sigma^j \sigma^{j+1} \sigma^j Q) \,= \,
A(P; \sigma^{j+1} \sigma^j \sigma^{j+1} Q)$
are fulfilled  identically. 
The Yang-Baxter equations with reflections 
arising from 
$ A(\tau^1\sigma^1\tau^1\sigma^1 P; Q ) \,=\,
    A(\sigma^1\tau^1\sigma^1\tau^1 P; Q ) $ 
and
$ A(\tau^N\sigma^{N-1}\tau^N\sigma^{N-1}P; Q ) \,=\,
    A(\sigma^{N-1}\tau^N\sigma^{N-1}\tau^NP; Q ) $
are satisfied identically because the reflection matrices
are just scalar factors.

As either $\{\sigma^1,\cdots,\sigma^{N-1},\tau^1\}$ or
$\{\sigma^1,\cdots,\sigma^{N-1},\tau^N\}$ can
be chosen as the basic elements of Weyl group
${\cal W}_B$, we must consider another consistency 
condition for $R_L$ and $R_R$. 
Using 
$\tau^1 = \sigma^1\sigma^2\cdots\sigma^{N-1}
           \tau^N\sigma^{N-1}\cdots\sigma^2\sigma^1$
and applying the $\check{S}$-matrices successively, 
we obtain an eigenvalue 
equation in the spinor space:
$\check{S}^{1,2}(\eta_1+\eta_2)
 \check{S}^{2,3}(\eta_1+\eta_3)\cdots 
 \check{S}^{N-1,N}(\eta_1+\eta_N) 
 \check{S}^{N-1,N}(\eta_1-\eta_N)\cdots 
 \check{S}^{2,3}(\eta_1-\eta_3)
 \check{S}^{1,2}(\eta_1-\eta_2) A(P;Q) 
 =R_L[(Pk)_1]R^{-1}_R[(Pk)_1]$ $ A(P;Q)$,
which is equivalent to 
\begin{eqnarray}
S^{1,2}(\eta_1 + \eta_2 )
S^{1,3}(\eta_1 + \eta_3 )\cdots 
S^{1,N}(\eta_1 + \eta_ N)\nonumber \\ 
S^{1,N}(\eta_1 - \eta_N)\cdots
%S^{1,3}(\eta_1 -\eta_3 )
S^{1,2}(\eta_1 - \eta_2 )A(P ; Q)\nonumber \\ 
= R_L[(Pk)_1] R^{-1}_R[(Pk)_1]A(P ; Q),
\label{eq:Eigenvalueproblem}
\end{eqnarray}
where $\eta_j = \sin (Pk)_j$.
This relation guarantees the consistency for any reflection
$(k_1,\cdots, \, k_j,\cdots )$ 
$\rightarrow (k_1,\cdots, \, -k_j,\cdots )$.

The eigenvalue problem (\ref{eq:Eigenvalueproblem})
can be diagonalized by means of Sklyanin approach 
\cite{Sklyanin}, in which the transfer matrix is
defined as 
$t(\alpha)=tr R^{-1}_L(\alpha)T_{+}(\alpha)
               R_R(\alpha)T_{-}(\alpha)$ 
where
$T_{\mp}(\alpha)=T_{AN}(\alpha \mp \alpha_N)\cdots
      T_{AN}(\alpha \mp \alpha_2) T_{AN}(\alpha \mp \alpha_1)$,
and 
$T_{Aj}(\alpha):=S^{Aj}(\alpha)\in End(V^A\otimes V^j)$.
The Bethe ansatz equations are obtained as follows
\begin{eqnarray}
 e^{-i2k_j L }= 
   e^{-i\theta_L( k_j )-i\theta_R( k_j )}
      \prod_{\nu=1}^{M}
          \Xi_{ \frac{1}{2} }(\eta_j -\lambda_{\nu} )
           \Xi_{ \frac{1}{2} }(\eta_j +\lambda_{\nu} ),
                \nonumber \\
 \prod_{\stackrel{\nu=1}{\nu\neq\mu} }^{M}
      \Xi_1 (\lambda_\mu -\lambda_{\nu} )
           \Xi_1^{-1}(\lambda_\mu +\lambda_{\nu} )
 =\prod_{l=1}^{N}
      \Xi_{ \frac{1}{2} }(\lambda_{\mu}-\eta_l )
          \Xi_{ \frac{1}{2} }(\lambda_{\mu} + \eta_l )
\nonumber 
\end{eqnarray}
where 
$\Xi_\beta(x) = (x-i\beta c)/(x+i\beta c)$. 
A set of coupled transcendental equations
are derived by taking the logarithm,
\begin{eqnarray}
 k_j = (\frac{\pi}{L})I_j 
         -\frac{1}{2L}[\theta_L(k_j) + \theta_R(k_j)]
           \hspace{12mm}\nonumber \\       
     +\frac{1}{2L}\sum_{\nu=1}^{M}
         \left[
            \Theta_{\frac{1}{2} }(\eta_j - \lambda_{\nu})
               +\Theta_{\frac{1}{2} }(\eta_j + \lambda_{\nu})
                 \right],
                    \nonumber \\
 \sum_{\nu=1}^{M}
       \left[
           \Theta_1 (\lambda_{\mu}- \lambda_{\nu})
              +\Theta_1(\lambda_{\mu} + \lambda_{\nu})
                 \right]
  =-2\pi J_\mu \nonumber \\             
     +\sum_{l=1}^{N}
        \left[
         \Theta_{\frac{1}{2} }(\lambda_{\mu}-\eta_l )
         +\Theta_{\frac{1}{2} }(\lambda_{\mu}+\eta_l )
           \right],
\label{eq:Secular}
\end{eqnarray}
where $\Theta_{\beta}(x) := 2\tan^{-1} (x/\beta c)$,
$-\pi < \tan^{-1}(x) \leq \pi $. 
The quantum numbers of both the quasi-charge and quasi-spin,
$I_j$ and $J_{\mu}$,
take integer values regardless of $N-M$ being even or odd. 
Eq.(\ref{eq:Secular}) becomes the result of Ref.\cite{Schulz}
in the limit $V_{L,R} \rightarrow \infty$.

We consider the thermodynamic limit by introducing 
the densities (distribution function)
of roots, $\rho (k)$ and $\sigma(\lambda)$ , and that of holes
$\rho_h(k)$ and $\sigma_h(\lambda)$. 
Then the secular equations (\ref{eq:Secular}) become
the following coupled integral equations
\begin{eqnarray}
 \rho (k) +\rho_h(k)
  &=&\frac{1}{ 2\pi }
       \left[ 2+\frac{1}{L}\frac{d}{dk}
          \left( \theta_L(k) + \theta_R(k)\right)
             \right]
             \nonumber \\          
  &+& \cos k\int^{B}_{-B} d\lambda'
           K_{\frac{1}{2} }(\sin k | \lambda')
            \sigma (\lambda'),
              \nonumber \\
 \sigma (\lambda)+ \sigma_h(\lambda) 
  &=& \int^{D}_{-D} dk'
        K_{\frac{1}{2} }(\lambda | \sin k')\rho (k')
          \nonumber \\ 
  &-& \int^{B}_{-B} d\lambda'
           K_{1}(\lambda |\lambda')\sigma  (\lambda') 
\label{eq:Density}
\end{eqnarray}
where $ K_{\beta}(x) :=\pi^{-1} \beta c /( \beta^2 c^2 + x^2 )$
and a notation $F(x|y) :=[ F(x-y) + F(x+y)]/2$ is used here and  
from now on.
The $B$ and $D$ are determined from the conditions 
\begin{equation}
\frac{1}{2}\int^{B}_{-B} d\lambda'\sigma (\lambda') 
   = \frac{M}{L}, \,\,\,\,
\frac{1}{2}\int^{D}_{-D} dk'\rho (k') = \frac{N}{L},\nonumber
\end{equation} 
where the factor $1/2$ is necessary \cite{footnote}
 
The ground state of the present model is a Fermi
sea described by
$\rho_{0}(k)$ and $\sigma_{0}(\lambda)$, where
$\rho_{0}(k)$ is the distribution function of charge 
with momentum $k$ and $\sigma_{0}(\lambda)$ that of 
down spins with respect to the rapidity $\lambda$. 
The distributions of the roots satisfy 
(\ref{eq:Density}) with $\rho_h=\delta(0)/L$, 
$\sigma_h=0$ and $B=\infty$.
%\begin{eqnarray}
% \rho_{0}(k) 
%  & = &  \frac{1}{ \pi }
%       \left[ 1+\frac{1}{2L}\frac{d}{dk}
%          \left(\theta_L(k) + \theta_R(k)\right)
%             \right]
%             \nonumber \\          
%  &\,& + \cos k\int^{B}_{-B} d\lambda'
%           K_{\frac{1}{2} }(\sin k | \lambda')
%            \sigma_{0}(\lambda'),
%              \nonumber \\
% \sigma_{0}(\lambda) 
%  &=& \int^{D}_{-D} dk'
%        K_{\frac{1}{2} }(\lambda | \sin k')\rho_{0}(k')
%          \nonumber \\ 
%  &\,& - \int^{B}_{-B} d\lambda'
%           K_{1}(\lambda |\lambda')\sigma_{0} (\lambda') 
%\label{eq:grounddensity}
%\end{eqnarray}
The energy of the ground state can be calculated once $\rho_{0}(k)$ 
is known. In this case, the distribution functions are solved
in a closed form by Fourier transform,
\begin{eqnarray}
2\pi\rho_{0}(k) &=&
    2 + \frac{1}{L}\frac{d}{dk}
          \left(\theta_L(k) + \theta_R(k)\,\right)
       - \frac{2\pi}{L}\delta(0)
             \nonumber \\   
   &+& \frac{2\pi}{c}\cos k  
        \int_{-D_0}^{D_0}d k' \rho_{0}(k')
           {\cal R}_{1}(\frac{\sin k'}{c} 
               \mid \frac{\sin k}{c}),
              \nonumber \\
\sigma_0 (\lambda ) &=&
 \frac{1}{8c}\int^{D_0}_{-D_0}dk\rho_0 (k)
   {\cal S}_1 \left(\frac{\pi}{c}\lambda 
                \mid \frac{\pi}{c}\sin k)\right),          
\label{eq:grounddensity2}
\end{eqnarray}
where we have used 
the following definitions,
\begin{eqnarray}
{\cal R}_{n}(x) = 
     \frac{1}{\pi} \sum^{\infty}_{l=1}
          (-n)^{l-1} \frac{l}{x^2 +l^2},
             \nonumber \\
{\cal S}_n (\frac{\pi}{2}x) =
      \frac{4}{\pi}\sum^{\infty}_{l=1}
          (- n)^{l+1} \frac{2l-1}{x^2 + (2l-1)^2}.
\nonumber
\end{eqnarray}                
Obviously, ${\cal S}_1(x)$ is the conventional hyperbolic
function sech$(x)$ which occurs in Lieb-Wu solution. 

Eqs.(\ref{eq:grounddensity2}) can be solved explicitly in some
special cases. In the strong coupling limit $ c \gg 1 $, 
it is simplified to
\begin{equation}
2\pi\rho_{str}(k)= 2 
      +\frac{2\pi}{L}\rho_b (k)
      -\frac{2\pi}{L}\delta(0)
      +\frac{4\pi}{c}
         (\frac{N}{L})\ln2\cos k,
\label{eq:Strong}
\end{equation}
where $2\pi\displaystyle\rho_b (k)= \frac{d}{dk}
        \left( \theta_L(k) + \theta_R(k)\right)$ 
arises from the boundary contribution 
at the ends of the wire.
The $D_0$ is completely determined 
from the explicit form (\ref{eq:Strong}),
hence the energy density of the ground state is obtained                                  
\begin{eqnarray}
\frac{E_0}{L}&=&
   -2t\left[\frac{1}{\pi}\sin(\frac{N\pi}{L})
            +\frac{\ln 2}{2c}\left(\frac{N}{L}\right)^2 
                ( 1 + \frac{\sin (2N\pi/L)}{2N\pi/L} )
                 \right]\nonumber \\
     &\,& +\frac{1}{L}(E_b + t) 
\label{eq:energy}
\end{eqnarray}
where
\begin{eqnarray}
 E_b =-\frac{t}{\pi}\left[\theta_L (\frac{N\pi}{2L}) +
           \theta_R (\frac{N\pi}{2L})
         \right]\cos(\frac{N\pi}{2L})
          \nonumber \\
   -\frac{t}{2\pi}\int^{N\pi/2L}_{-N\pi/2L}
        [\theta_L (k) + \theta_R (k)]\sin k dk,
\nonumber 
\end{eqnarray}
which vanishes when $V_{L,R} \rightarrow \infty $.
It is worthwhile to compare (\ref{eq:energy}) with the result
in the case of periodic boundary condition. The difference is 
calculated,
\[
  \frac{E_0 - E_0^{per} }{L} = 
           \frac{1}{L}(E_b + t)
\]
where $E^{per}_0$ stands for the ground-state  energy with 
periodic boundary condition. Clearly, the difference is
relatively small for a large system which agrees with the
common understanding about the effects of boundary condition.

It is convenient to study the excitations by introducing
$\rho(k)=\rho_0(k) +\rho_1(k)/L$ and
$\sigma(\lambda)=\sigma_0 (\lambda)+\sigma_1(\lambda)/L$
where $\rho_0(k)$ and $\sigma_0 (\lambda)$ satisfying
(\ref{eq:Density}) with $\rho_h(k) =\delta(0)/L$ 
and $\sigma_h (\lambda)=0$.
The excited energy up to the order $O(1/L)$ is
\begin{equation}
 \Delta E = -\int_{-D}^D dk (2t\cos k + \mu)\rho_1 (k),
\label{eq:Excite}
\end{equation}
where $\mu$ stands for the chemical potential \cite{Cornelius}.
The (\ref{eq:Excite}) is related to $\rho_1(k)$ only
and is valid for both the spin excitation and the charge
excitation. For a large system the $D$ can be replaced by
$D_0$.

The simplest spin excitation is a triplet,
i.e., two-hole state with
$\sigma_h (\lambda)=[\delta (\lambda\mid\bar{\lambda}_1)
     +\delta(\lambda\mid\bar{\lambda}_2)]/L$. 
The excitation energy is composed of two terms
$\Delta E_{trip} = \varepsilon_s (\bar{\lambda}_1)
            +\varepsilon_s (\bar{\lambda}_2)$,
each of them can be identified as a spinon excitation energy
\begin{equation}
\varepsilon_s (\bar{\lambda})=
  -\int_{-D_0}^{D_0}dk (2t\cos k + \mu )
     \rho^s_1 (k, \bar{\lambda}),
\label{eq:Tripenergy}
\end{equation}
where $\rho^s_1 (k, \bar{\lambda})$ solves the following
integral equation,
\begin{eqnarray}
\rho^s_1 (k;\bar{\lambda})=
  -\frac{\cos k}{8c}{\cal S}_1
    \left(\frac{\pi}{c}\bar{\lambda}\mid\frac{\pi}{c}\sin k
     \right) \hspace{12mm} \nonumber \\
 + \frac{2\cos k}{c} \int^{D_0}_{-D_0}dk'{\cal R}_1
      \left(\frac{\sin k}{c} \mid\frac{\sin k'}{c}
       \right)\rho^s_1 (k', \bar{\lambda}).
\label{eq:Sdensity}
\end{eqnarray}
From the asymptotic behavior of the ${\cal S}_1$
one can find that the spinon excitation is gapless.

The excitation in charge configuration is a variation
from the ground state in the sequence of charge 
quantum numbers $\{I_j\}$. The simplest case is the
holon-antiholon excitation, i.e.,
one hole $k_h \in [-D_0, D_0]$  and one `particle'
$k_p \notin [-D_0, D_0]$  state. In this case
$\rho_h =\delta (k\mid k_h)/L$, the integral equations
(\ref{eq:Density}) give
\begin{eqnarray}
 \rho^c_1 (k) &=&
      - \delta (k\mid k_h) 
       + \cos k\int^{\infty}_{-\infty} d\lambda'
           K_{\frac{1}{2}}(\sin k | \lambda')
            \sigma^c_1 (\lambda'),
              \nonumber \\
 \sigma^c_1 (\lambda)
  &= & K_{\frac{1}{2}}(\lambda |\sin k_p)
       + \int^{D_0}_{-D_0} dk'
          K_{\frac{1}{2} }(\lambda | \sin k')\rho^c_1(k')
           \nonumber \\ 
  &\,& -\int^{\infty}_{-\infty}d\lambda'
           K_{1}(\lambda |\lambda')\sigma^c_1 (\lambda'). 
\label{eq:Cdensity}
\end{eqnarray}
The excitation energy is composed of two terms
$\Delta E = \varepsilon_c (k_h)
             - \varepsilon_c (k_p)$.
They can be identified as a holon excitation energy 
$\varepsilon_c (k_h)$ and an antiholon (particle state)
excitation energy 
$\bar{\varepsilon}_c = -\varepsilon_c (k_p)$
respectively, namely,
\begin{equation}
\varepsilon_c (\bar{k})= 2t\cos\bar{k}
    -\int_{-D_0}^{D_0}dk
       (2t\cos k + \mu)\rho^c_1(k, \bar{k}),
\end{equation}
where the $\rho^c_1(k, \bar{k})$ is determined by 
\begin{eqnarray}
\rho^c_1 (k; \bar{k}) &=& \frac{2\cos k}{c} 
            \left[ {\cal R}_1\left(\frac{\sin k}{c}
             \mid\frac{\sin\bar{k}}{c}\right)\right.
              \nonumber\\
 &\,& \left.
       +\int_{-D_0}^{D_0}dk' {\cal R}_1
         \left(\frac{\sin k}{c}
          \mid \frac{\sin k'}{c}\right)
           \rho^c_1 (k'; \bar{k}) \right].
            \nonumber
\end{eqnarray}
Since both $k_h$ and $k_p$ can tend to $D_0$,
the holon-antiholon excitation is of no gap. 
Further case is the holon-holon excitation which involves complex 
$k$-pairs and is gapful.

The excitation in spin configuration is allowed to form a singlet,
i.e., two holes and a $2$-string 
$\bar{\lambda}^{\pm}=(\bar{\lambda}_1+\bar{\lambda}_2)/2          
 \pm ic/2$.
The excitation energy is composed of two terms
\begin{equation}
\Delta E_{sing}= 
       \varepsilon_s (\bar{\lambda}_1)
       +\varepsilon_s (\bar{\lambda}_2).
\end{equation}
These two terms are free spinon excitation energies given 
by (\ref{eq:Tripenergy}-\ref{eq:Sdensity}).

In summary, we obtained an exact solution for the Hubbard model
with the realistic quantum-wire boundary condition that the 
possibility of a nonvanishing wave function outside is taken
into account.   
The ground state and some low-lying excitations are studied
in the thermodynamic limit. 
            
YQL acknowledges the grands
NSFC-19675030 and NSFZ-194037, support from 
the Y. Pao \& Z. Pao Foundation, 
as well as the discussions with Dr. D. F. Wang 
and Prof. F. C. Zhang.

\end{multicols}


\begin{thebibliography}{99}
\bibitem{footnote} 
There are symmetries in the
solution space of the roots $\{ k_j\}$ and $\{\lambda_{\mu}\}$.
For instance, such a transformation from solution
$(\lambda_1, \cdots,\lambda_{\mu},\cdots,\lambda_M)$
to solution
$(\lambda_1,\cdots,-\lambda_{\mu},\cdots,\lambda_M)$
keeps the roots $\{k_j\}$  unchanged that the two points
with $\lambda_{\mu}$ and $-\lambda_{\mu}$ label the same
state etc. Consequently, we must discount the twice counting
when taking thermodynamic limit in the usual way (i.e.
taking integrals from $-D$ to $D$ for $k$ and from
$-B$ to $B$ for $\lambda$). We also must minus the point
of $k=0$, which is realized by placing a hole at the origin
of $k$.
\end{thebibliography}
\end{document}